\documentclass[aps%
 ,twocolumn%
 ,secnumarabic%
,amssymb, amsmath,nobibnotes, aps, pre, floatfix,superscriptaddress]{revtex4-1}

\usepackage{graphicx,epsfig}
\usepackage{amsmath}
\usepackage{amsfonts}
\usepackage{fancyhdr}
\usepackage{hyperref}
\usepackage{epstopdf}
\usepackage{longtable}
 \usepackage{comment}
 \usepackage{enumitem}  

\begin{document}

\pagestyle{fancy}
\lhead{}
\rhead{V. Navas-Portella, \'A. Gonz\'alez, I. Serra, E. Vives and \'A. Corral}
\lfoot{}
\rfoot{}
\title{Universality of power-law exponents by means of maximum likelihood estimation}
\author{V\'ictor Navas-Portella}
\affiliation{Centre de Recerca Matem\`atica, Edifici C, Campus Bellaterra, E-08193 Bellaterra, Catalonia, Spain.}
\affiliation{Barcelona Graduate School of Mathematics, Edifici C, Campus Bellaterra,
E-08193 Barcelona, Spain.} 
\affiliation{Facultat de Matem\`atiques i Inform\`atica, Universitat de Barcelona, Barcelona, Spain.} 

\author{\'Alvaro Gonz\'alez}
\affiliation{Centre de Recerca Matem\`atica, Edifici C, Campus Bellaterra, E-08193 Bellaterra, Catalonia, Spain.}

\author{Isabel Serra}
\affiliation{Centre de Recerca Matem\`atica, Edifici C, Campus Bellaterra, E-08193 Bellaterra, Catalonia, Spain.}

\author{Eduard Vives}
\affiliation{Departament de Mat\`eria Condensada, Facultat de F\'{\i}sica, Universitat de Barcelona, Mart\'i Franqu\`es 1, 08028 Barcelona, Catalonia, Spain.}
\affiliation{Universitat de Barcelona Institute of Complex Systems (UBICS), Facultat de F\'{\i}sica, Universitat de Barcelona, Barcelona, Catalonia, Spain.}

\author{\'Alvaro Corral}
\affiliation{Centre de Recerca Matem\`atica, Edifici C, Campus Bellaterra, E-08193 Bellaterra, Catalonia, Spain.}
\affiliation{Barcelona Graduate School of Mathematics, Edifici C, Campus Bellaterra,
E-08193 Barcelona, Spain.} 
\affiliation{Complexity Science Hub Vienna,
Josefst\"adter Stra$\beta$e 39,
1080 Vienna,
Austria.}
\affiliation{Departament de Matem\`atiques, Universitat Aut\`onoma de Barcelona,
E-08193 Barcelona, Spain.}

\begin{abstract}
Power-law type distributions are extensively found when studying the behaviour of many complex systems. However, due to limitations in data acquisition, empirical datasets often only cover a narrow range of observation, making it difficult to establish power-law behaviour unambiguously. 
In this work we present a statistical procedure to merge different datasets with the aim of obtaining a broader fitting range for the statistics of different experiments or observations of the same system or the same universality class.
This procedure is applied to the Gutenberg-Richter law for earthquakes and for synthetic earthquakes (acoustic emission events) generated in the laboratory: labquakes. Different earthquake catalogs have been merged finding a Gutenberg-Ricther law holding for more than eight orders of magnitude in seismic moment. The value of the exponent of the energy distribution of labquakes depends on the material used in the compression experiments. By means of the procedure exposed in this manuscript, it has been found that the Gutenberg-Richter law for earthquakes and charcoal labquakes can be characterized by the same power-law exponent.
\end{abstract}

\maketitle

\section{Introduction}
\label{intro}


Generally speaking, a complex system can be understood as a large number of interacting elements whose global behaviour cannot be derived from the local laws that characterize each of its components. The global response of the system can be observed at different scales and the vast number of degrees of freedom makes prediction very difficult. In this context, a probabilistic description of the phenomenon is needed in order to reasonably characterize it in terms of random variables. When the response of these systems exhibits lack of characteristic scales, it can be described in terms of power-law type probability density functions (PDF), $f(x) \propto x^{-\gamma}$, where $x$ corresponds to the values that the random variable that characterizes the response of the system can take,  $\propto$ denotes proportionality and the power-law exponent $\gamma$ acquires values larger than one. The power-law is the only function which is invariant under any scale transformation of the variable $x$ \cite{Christensen_Moloney}. This property of scale invariance confers a description of the response of the system where there are no characteristic scales. This common framework is very usual in different disciplines \cite{bak1997nature,Sethna2001,Corral2018} such as condensed matter physics \cite{saljedahmen}, geophysics \cite{Corral2018}, seismology \cite{kagan2013earthquakes,Kaganbeta}, economics \cite{Gabaix2016}, linguistics \cite{moreno2016}, etc. 


It has been broadly studied \cite{StanleyRMP1999,STANLEY200060} that different complex systems present common values of all their power-law exponents and can be grouped into the same universality class. Therefore, it is important to determine rigorously these exponents, not only to properly characterize phenomena but also to provide a good classification into universality classes.

In practice, exponents are difficult to measure empirically. Due to experimental limitations that distort the power-law behaviour, the property of scale invariance can only be measured in a limited range. Therefore, when a power-law distribution is fitted to empirical data is more convenient to talk about local or restricted scale invariance. In this context, the wider the range the fitted PDF spans, the more reliable and strong this property will be.

A paradigmatic example of power-law behaviour in complex systems is the well-known Gutenberg-Richter (GR) law for earthquakes \cite{Utsu1999}.
This law states that, above a lower cut-off value, 
earthquake magnitudes follow an exponential distribution;
in terms of the magnitude PDF
\begin{equation}
f(m)  = (b \ln 10) 10^{-b(m-m_{min})}
\propto 10^{-b m},
\end{equation}
defined for $m\ge m_{min}$,
with $m$ the magnitude (moment magnitude in our case),
$m_{min}$ the lower cut-off in magnitude and 
$b$ the so called $b-$value.
The general relationship between seismic moment $x$ and moment magnitude $m$ is given by:
\begin{equation}
x = 10^{\frac{3}{2}m +9.1}
\label{eq:moments}
\end{equation}
measured in units of Nm \cite{Kanamori79}.
The GR law is a power-law distribution when it is written as a function of the seismic moment $x$:
\begin{equation}
f(x) = \frac{2}{3} \frac{b}{x_{min}} \left(  \frac{x}{x_{min}} \right)^{-\left( 1+\frac{2}{3}b \right)} = \frac{\gamma -1}{x_{min}} \left( \frac{x}{x_{min}} \right)^{-\gamma}
\label{eq:notrunc}
\end{equation}
where we conveniently define $\gamma=1+\frac{2}{3}b$ and $x_{min}$ corresponds to the value of the seismic moment of the cut-off magnitude $m_{min}$ \cite{Kanamori79} introduced in Eq.~\ref{eq:moments}. Note that this PDF has a domain $x \in \left[ x_{min}, +\infty \right)$ as $b>0$, then $\gamma>1$.

An earthquake catalog is an empirical dataset that characterizes each earthquake by an array of observations: time of occurrence, spatial coordinates, magnitude, etc.
The magnitude $m_{min}$ is usually regarded as the completeness threshold, such as all earthquakes with $m \geq m_{min}$ are recorded in the catalog \cite{Woessner2005}. For $m<m_{min}$, events are missing from the catalog due to the difficulties of detecting them (e.g.\cite{Gonzalez2017,Schorlemmer2008}), specially in aftershock sequences, when their waveforms tend to overlap each other \cite{Woessner2005,Davidsen2010}. This incompleteness distorts the power-law behaviour below $m_{min}$.
  One has also to keep in mind that there also exists an upper cut-off due to the finite-size effects \cite{CorralRosalba2018}, implying that, at a certain value of the magnitude, there are deviations from the power-law behaviour. Consequently, strictly speaking, the range of validity of the GR law cannot be extended up to infinity \cite{Serra2017,Corral2018}. 
By ignoring which is the model that conveniently fits the tail of the distribution, the power-law behaviour has to be restricted to an upper cut-off $x_{max}$ and the PDF for the truncated GR law is written:
\begin{equation}
f(x) = \frac{1-\gamma }{x_{max}^{1-\gamma}-x_{min}^{1-\gamma}} x ^{-\gamma},
\label{eq:trunc}
\end{equation}
defined for $x \in \left[ x_{min}, x_{max} \right]$.

Recent studies regarding the acoustic emission (AE) in compression experiments of porous glasses and minerals \cite{Baro2013,Nataf2014,Navas-Portella2016,Yoshimitsu2014,Main2013} or wood \cite{Makinen2015} have focused the attention in the energy 
distribution of AE events due to the similarities with the GR law for earthquakes \cite{Serra2017}. According to the terminology which is used in some of these studies, we will name as labquakes those AE events that occur during the compression of materials.

Earthquake and labquake catalogs as well as other empirical datasets in complex systems only report a limited range of events, making it difficult to estimate parameters of the power-law PDF accurately. In this work we try to solve this problem by combining datasets with rigorous statistical tools, with the goal of finding a broader range of validity. In addition, if different datasets can be combined and characterized by a unique power-law exponent, that means that the particular exponents of each dataset are statistically compatible, therefore, providing a possible way of classifying them into  the same universality class.

The manuscript is structured as follows: in Sec. \ref{sec:methods} we will expose the different statistical procedures that are used in order to conveniently merge datasets and to find a global power-law PDF (\ref{sec:statistic}). In Sec. \ref{sec:applications} we apply this methodology to different datasets that are explained Sec.\ref{sec:datasets} and analysed in Sec. \ref{sec:earthquake} for earthquake catalogs and in Secs.\ref{sec:earthquake+charcoal} and \ref{sec:vycor} for AE data obtained during the compression of two different materials.

\section{Methods}
\label{sec:methods}
\subsection{Statistical methodology: Merging datasets }

 \label{sec:statistic}
By considering $n_{\rm ds}$ datasets of $N_{i}$ ($i=1,...,n_{\rm ds}$) observations each, one wants to fit a general power-law distribution with a unique global exponent for all of them. We assume that for the $i$-th dataset, the variable $\mathcal{X}$  (seismic moment if one works with the GR law for earthquakes or AE energy if one works with the GR law for labquakes) follows a power-law PDF $f_{\mathcal{X}}(x; \gamma_{i},x_{min}^{(i)},x_{max}^{(i)})$ from a certain lower cut-off $x_{min}^{(i)}$ to an upper cutt-off $x_{max}^{(i)}$ with exponent $\gamma_{i}$ and number of data $n_{i}$ ($n_i \leq N_i$) in the range $\left[ x_{min}^{(i)},x_{max}^{(i)}\right]$. Note that one also can consider the untruncated power-law model for the $i$-th dataset if $x_{max}^{(i)}\rightarrow \infty$. By means of the methods explained in Refs. \cite{Deluca2013,Corral2018} (or, alternatively, Ref.~\cite{Clauset2009}), one can state that data from the $i$-th dataset does not reject the power-law hypothesis for a certain range. Note that, in  the $i$-th dataset, the variable $\mathcal{X}$ can acquire values in a range typically spanning several orders of magnitude. 

Generally, the procedure of merging datasets is performed by selecting upper and lower cut-offs $x_{min}^{(i)}$ and $x_{max}^{(i)}$ ($x_{min}^{(i)} < x_{max}^{(i)}$) for each dataset. Data outside these ranges are not considered. All the possible combinations of cut-offs $\lbrace x_{min} \rbrace$ and $\lbrace x_{max} \rbrace$ are checked with a fixed resolution (see below). The Residual Coefficient of Variation (CV) test can be used to fix some upper cut-offs, thus reducing the computational effort. For more details about the CV test, see Appendix \ref{sec:appa}.

Given a set of cut-offs, datasets can be merged by considering two different models:
\begin{itemize}
\item \textbf{Model $\alpha$}: All datasets are merged by considering a unique global exponent $\Gamma$ ($\gamma_i=\Gamma$ for all datasets). 
\item \textbf{Model $\beta$}: All datasets are merged, but each one with its own exponent $\gamma_i$ ($i=1,...,n_{\rm ds}$).
\end{itemize}
Note that model $\alpha$ is nested in model $\beta$ and the difference in the number of parameters characterizing these models is $n_{\beta}-n_{\alpha}=n_{ds}-1$. Since we are interested in merging datasets with a unique global exponent (model $\alpha$), we need enough statistical evidence that this simpler model is suitable to fit the data.
For a given set of cut-offs, the fit is performed by means of the following protocol:
\begin{enumerate}
\item \textbf{Maximum Likelihood Estimation (MLE) of model $\alpha$}: The log-likelihood function of model $\alpha$ can be written as:
\begin{equation}
\log \mathcal{L}_{\alpha}= \sum_{i=1}^{n_{\rm ds}} \sum_{j=1}^{n_{i}} \log f_{\mathcal{X}} \left( x_{ij}; \Gamma, x_{min}^{(i)},x_{max}^{(i)} \right)
\label{eq:general}
\end{equation}
 where $x_{ij}$ corresponds to the $n_i$ values of the variable $\mathcal{X}$ that are in the range $x_{min}^{(i)} \leq x_{ij} \leq x_{max}^{(i)}$ in the $i$-th dataset, $\log$ is the natural logarithm and $\Gamma$ is the global exponent. The definition of likelihood is consistent with the fact that likelihoods from different datasets can be combined in this way \cite{pawitan2013all}[p.27]. At this step, one has to find the value $\hat{\Gamma}$ of the global exponent $\Gamma$ that maximizes the log-likelihood expression in Eq. (\ref{eq:general}). For the particular expressions corresponding to the truncated and untruncated power-law PDF, see Eqs. (\ref{eq:likepl}) and (\ref{eq:likepltrunc}) in Appendix \ref{sec:appb}. If all the power-law distributions are untruncated, this exponent can be easily found analytically \cite{Navas2018} as $$\hat{\Gamma} = 1 + \frac{ \sum_{i=1}^{n_{ds}} n_{i}}{\sum_{i=1}^{n_{ds}} \frac{n_{i}}{\hat{\gamma}_{i}-1}} $$
where the hats denote the values of the exponents that maximize the log-likelihood of the particular power-law distribution (model $\beta$) and the general one in Eq.(\ref{eq:general}). 
  If truncated power-law distributions are considered, one has to use a numerical method in order to determine the exponent $\Gamma$ that maximizes this expression \cite{Baro2012}.  
\item \textbf{MLE of model $\beta$}: The log-likelihood function of model $\beta$ can be written as:
\begin{equation}
\log \mathcal{L}_{\beta}= \sum_{i=1}^{n_{\rm ds}} \sum_{j=1}^{n_{i}} \log f_{\mathcal{X}} \left( x_{ij}; \gamma_{i}, x_{min}^{(i)},x_{max}^{(i)} \right)
\label{eq:generalbeta}
\end{equation}
 using the same notation as in Eq. (\ref{eq:general}). For the particular expressions corresponding to the truncated and untruncated power-law PDFs, see Eqs. (\ref{eq:likepl}) and (\ref{eq:likepltrunc}) in Appendix \ref{sec:appb}. The values of the exponents that maximize Eq.(\ref{eq:generalbeta}) are denoted as $\hat{\gamma}_i$.
 
\item \textbf{Likelihood-ratio test}: We perform the Likelihood Ratio Test (LRT) for the models $\alpha$ and $\beta$ in order to check whether model $\alpha$ is good enough to fit data or not in comparison with model $\beta$. For more details about the LRT see Appendix \ref{sec:appb}. If model $\alpha$ ``wins'', go to step (4). Otherwise, this fit is discarded and a different set of cut-offs $\lbrace x_{min} \rbrace$ and $\lbrace x_{max} \rbrace$ is chosen, and go back to step (1). Note that model $\alpha$ can be a good model to fit if the particular exponents $\hat{\gamma}_i$ do not exhibit large differences among each other in relation to their uncertainty. 

\item \textbf{Goodness-of-fit test}: In order to check whether it is reasonable to consider model $\alpha$ as a good candidate to fit data, we formulate the next null hypothesis $\rm H_{0}$: the variable $\mathcal{X}$ is power-law distributed with the global exponent $\hat{\Gamma}$ for all the datasets. In this work, we are going to use two different statistics in order to carry out the goodness-of-fit tests: the Kolmogorov-Smirnov Distance of the Merged Datasets (KSDMD) and the Composite Kolmogorov-Smirnov Distance (CKSD). The KSDMD statistic can be used as long as datasets overlap each other whereas the CKSD statistic does not require this condition. For more details about how these statistics are defined and how the $p$-value of the test is found see Appendix \ref{sec:appd}.  If the resulting $p$-value is greater than a threshold value $p_{c}$ (in the present work we are going to use $p_{c}=0.05$ and $p_{c}=0.20$), we consider this as a valid fit and it can be stated that the variable $\mathcal{X}$ is power-law distributed with exponent $\hat{\Gamma}$ along all the different datasets for the different ranges $\lbrace x_{min} \rbrace$ and $\lbrace x_{max} \rbrace$. Otherwise, this fit will not be considered as valid. A different set of cut-offs is chosen and we go back to step (1).

\end{enumerate}

When all the combinations of cut-offs have been checked, one may have a list of valid fits. In order to determine which one of them is considered the definitive fit, the following procedure is carried out:

\begin{enumerate}[label=(\roman*)]
\item The fit that covers the largest sum of orders of magnitude: 
$ \max \left[ \sum_{i=1}^{n_{ds}} \log_{10}\left(\frac{x_{max}^{(i)}}{x_{min}^{(i)}} \right) \right] $ is chosen. If the power-law fit is untruncated, $x_{max}^{(i)}$ can be substituted by the maximum observed value  $x_{top}^{(i)}$. If there is a unique candidate with a maximum number of orders of magnitude, then this is considered as the definitive global fit. Otherwise go to the next step (ii).
\item The fit with the broadest global range: $\max \left[ \frac{\max \left( x_{max}^{(i)}  \right)}{ \min \left( x_{min}^{(i)} \right)  } \right]$ for $i=1,...,n_{ds}$ is chosen. If there is a unique candidate, this is considered as the definitive global fit. Otherwise go to the next step (iii).
\item The fit with the maximum number of data $\mathcal{N}=\sum_{i=1}^{n_{ds}}n_i$ is considered as the definitive global fit.
\end{enumerate}

By means of these three steps, a unique fit has been found for all the datasets analysed in this work. Nevertheless, one could deal with datasets in which more conditions are needed in order to choose a definitive fit unambiguously. At the end of this procedure one is able to state that the datasets that conform the global fit correspond to phenomena that are candidates to be be classified into the same universality class, at least regarding the observable $\mathcal{X}$. If no combination of cut-offs is found to give a good fit, then it can be said that there exists at least one catalog that corresponds to a phenomenon that must be classified in a different universality class. 


\section{Applications}
\label{sec:applications}
In this section we apply the methodology of merging datasets to different earthquake and labquake catalogs. Firstly, we try to merge three earthquake catalogs. Secondly, a fourth  catalog of charcoal labquakes is added in order to check whether these two phenomena can be classified into the same universality class. Finally, we apply the methodology to four catalogs of Vycor labquakes that cover different observation windows. 
\subsection{Datasets}

\label{sec:datasets}
\subsubsection{Earthquake Catalogs}
\begin{figure}
 \centering
  \includegraphics[scale=0.35]{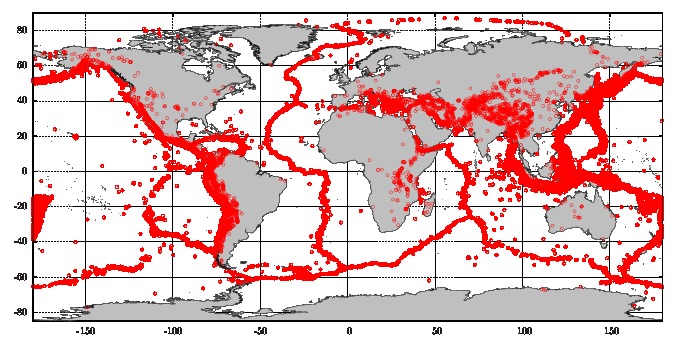}
  \includegraphics[scale=0.45]{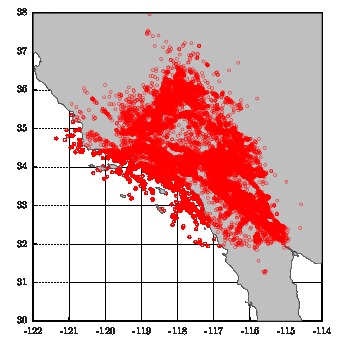}
  \includegraphics[scale=0.45]{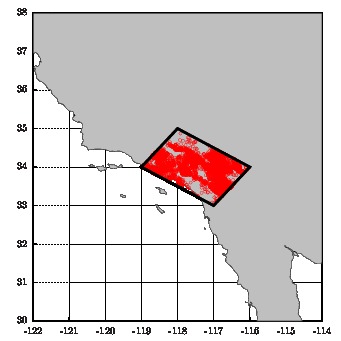}
 \caption{Earthquake epicenters of the different catalogs. Horizontal and vertical axes correspond to longitude and latitude respectively in degrees. Top, CMT catalog for the period 1977-2017 \cite{CMT1,CMT2}. Middle, YHS A Catalog \cite{2012BuSSA.102.1179Y} of Southern California for the period 1981-2010. Bottom figure corresponds to a region (YHS B) of the YHS catalog for Los Angeles area for the period 2000-2010.} \label{fig:mapzoom} 
\end{figure}

We have selected catalogs that have different completeness magnitudes in order to cover different magnitude ranges.
We hope that a convenient combination of these datasets will give us a larger range of validity of the GR law.
Let us briefly describe the catalogs that have been used in this work (See also Fig.~\ref{fig:mapzoom}):
\begin{itemize}
\item Global Centroid Moment Tensor (CMT) Catalog: It comprehends earthquakes worldwide from 1977 to 2017 \cite{CMT1,CMT2}. This catalog reports the values of the moment magnitude as well as the seismic moment. The resolution of the magnitude in the catalog is $\Delta m=0.01$. 
\item Yang--Hauksson--Shearer (YHS) A: It records the earthquakes in Southern California with $m \geq 0$ in the period 1981-2010 \cite{2012BuSSA.102.1179Y}. This catalog does not report the seismic moment but a preferred magnitude that is converted into seismic moment according to Eq.(\ref{eq:moments}). The resolution of the catalog is $\Delta m=0.01$. 
\item Yang--Hauksson--Shearer (YHS) B: It is a subset of YHS A that contains the earthquakes in the region of Los Angeles in the period 2000-2010 \cite{2012BuSSA.102.1179Y}. LA region is defined by the following four vertices in longitude and latitude: $\left(119^{\circ}W, 34^{\circ}N \right)$,$\left(118^{\circ}W, 35^{\circ}N \right)$,$\left(116^{\circ}W, 34^{\circ}N \right)$, and $\left(117^{\circ}W, 33^{\circ}N \right)$ ( see Fig.~\ref{fig:mapzoom}). This region has been selected because it is among the best monitorized ones \cite{Hutton,Schorlemmer2008}. Furthermore, we selected this time period due to the better detection of smaller earthquakes than in previous years \citep{Hutton} what should reduce the completeness magnitude of the catalog \cite{Schorlemmer2008}. The resolution of this catalog is the same as YHS A.
\end{itemize}
In order to not to count the same earthquake more than once, the spatio-temporal window corresponding to the YHS B catalog has been excluded from the YHS A and the spatio-temporal window corresponding to the YSH A catalog has been excluded from the CMT catalog. Figure \ref{fig:mapzoom} shows the epicentral locations of the earthquakes contained in each catalog.

\subsubsection{Labquake Catalogs}

We performed uni-axial compression experiments of porous materials in a conventional test machine ZMART.PRO (Zwick/Roell). Samples with no lateral confinement were placed between two plates that approached each other at a certain constant displacement rate $\dot{z}$. Simultaneous to the compression, recording of an AE signal was performed by using a piezoelectric transducer embedded in one of the compression plates. The electric signal $U(t)$ was pre-amplified, band filtered (between 20 kHz and 2 MHz), and analysed by means of a PCI-2 acquisition system from Euro Physical Acoustics (Mistras Group) with an AD card working at 40 Megasamples per second with 18 bits precision \cite{PCI2}. Signal pre-amplification was necessary to record small AE events. Some values of the pre-amplified signal were so large that cannot be detected correctly by the acquisition system. This fact led to signal saturation and, consequently, an underestimated energy of the AE event \cite{Navas2018}. Recording of data stopped when a big failure event occurred and the sample got destroyed.

An AE event (often called AE hit in specialized AE literature) starts at the time $t_{j}$ when the signal $U(t)$ exceeds a fixed detection threshold and finishes at time $t_{j}+\tau_{j}$ when the signal remains below threshold from $t_{j}+\tau_{j}$ to at least $t_{j}+\tau_{j}+200\mu$s. 
The energy $E_{j}$ of each event is determined as $E_{j}=\frac{1}{R}\int_{t_{j}}^{t_{j}+\tau} U^{2}(t) dt$ where $R$ is a reference resistance of $10$ k$\Omega$. This AE energy corresponds to the radiated energy received by the transducer. At the end of an experiment, a catalog of events is collected, each of them characterized by a time of occurrence $t$, energy $E$, and duration $\tau$.

Experiments were performed with two different materials:
\begin{itemize}
\item \textbf{Charcoal}: One experiment with a charcoal sample was performed at a constant rate $\dot{z} = 0.005$mm/min  with a preamplification of $40$ dB and a value of $43$dB for the detection threshold. The sample corresponded to commercially available fine art fusains (HB5 mm, NITRAM, Canada).

\item \textbf{Vycor}: We use the labquake catalogs of Ref.~\cite{Navas2018}, in which the authors performed four experiments with cylindrical samples (diameters $\Phi = 4.45$ mm and heights $H=8$ mm) of Vycor (a mesoporous silica glass with $40 \%$ porosity) were performed at a constant rate $\dot{z}=0.005$mm/min. Before compression, samples were cleaned with a $30\%$ solution of $\rm H_{2}O_{2}$, during 24 h and dried at 130$^{\circ}$C.
The four experiments were performed with the following pre-amplification values: $60$ dB, $40$ dB, $20$ dB and $0$ dB, and the respective values of the detection threshold $23$ dB, $43$ dB, $63$ dB and $83$ dB referring to the signal $U(t)$, not the preamplified signal (in such a way that after preamplification the threshold always moves to $83$ dB). This value of the threshold was set as low as possible in order to avoid parasitic noise. 

\end{itemize}

An important difference between these two materials is the degree of heterogeneity. The mesoporous silica structure of Vycor is much more homogeneous than the one of the charcoal which may contain voids and macropores. These structural differences may lead to differences in the energy exponents \cite{Vives2019}.

\subsection{Merging Earthquake Catalogs}
\label{sec:earthquake}
\begin{table*}[htbp]
\begin{tabular}{|ll|r|r|r|r|r|r|l|l|l|}
\hline
 &   &   \multicolumn{1}{l|}{\textbf{$N$}} & \multicolumn{1}{l|}{\textbf{$n$}} & \multicolumn{1}{l|}{\textbf{$m_{min} $}} &  \multicolumn{1}{l|}{\textbf{$x_{min}$ (Nm)}} & \multicolumn{1}{l|}{\textbf{$m_{top}$}} & \multicolumn{1}{l|}{\textbf{$x_{top}$(Nm)}} &  \multicolumn{1}{l|}{\textbf{$\hat{b}$-value}}  &   \multicolumn{1}{l|}{\textbf{$\hat{\gamma}$}} & \multicolumn{1}{l|}{\textbf{$p_{fit}$}} \\ \hline
& \textbf{(0) Charcoal} & 101524 & 18625 & - & $4.93\times 10^{-18}$ & - & $1.86 \times 10^{-11}$ & $0.984(8)$ & $1.656(5)$  &  $0.088(9)$ \\ \hline 
& \textbf{(1) YHS B}  & 26330 & 3412 & 1.93  & $10^{12}$& 5.39 & $1.53 \times 10^{17}$  & $0.99(3)$  &  $1.66(2) $  & $0.072(8)$   \\ \hline 
& \textbf{(2) YHS A} & 152924 & 4353 & 3.17 & $7.08 \times 10^{13}$ & 7.20 & $7.94 \times 10^{19}$  & $0.98(1)$ &  $1.65(1)$  & $0.080(9)$ \\ \hline
& \textbf{(3) CMT} & 48637 &  22336 & 5.33 &  $1.24 \times 10^{17}$ & 9.08 & $5.25 \times 10^{22}$  & $0.982(7)$ &  $1.655(5)$ & $0.36(2)$  \\ \hline
  
\end{tabular}
\caption{Results of fitting the GR law for each individual catalog. The total number of earthquakes in each catalog is given by $N$ whereas the number of data entering into the fit is $n$. The value $m_{top}$ corresponds to the maximum observed value for each catalog. The GR law is valid for each catalog from $\left[m_{min}, m_{max} \right]$ with a particular $b$-value. $m_{max}$ has no upper-limit for any of the fits except for the CMT catalog, in which $m_{max}^{(3)}=7.67$ ($x_{max}^{(3)}=4.03 \times 10^{20}$ Nm). For the charcoal catalog, $x$ represents the energy collected by the transducer (instead of the seismic moment). No magnitude scale is provided in this case. Numbers in parentheses correspond to the error bar estimated with one $\sigma$ in the scale given by the last digit. The $p$-value of the fits has been computed with $10^3$ simulations and $p_c=0.05$. For completeness, values of cut-offs in seismic moment $x$ and power-law exponents $ \hat{\gamma}$ have been included.}
  \label{tab:universaltable}
\end{table*}

In its usual form, the GR law fits a power-law model that contemplates a unique power-law exponent. Nevertheless, several studies have elucidated the existence of a double-power-law behaviour in the GR law for global seismicity \cite{Corral2018,Pacheco1992,YODER2012167}. Authors in Ref.~\cite{Corral2018} claim that a truncated power-law with exponent $\gamma \simeq 1.66$ cannot be rejected up to $m_{max} \simeq 7.4$ and a second power-law tail emerges from $m_{min}'=7.67$ with an exponent $\gamma'= 2.1\pm 0.1$.

We have checked these results by using the residual CV-test. Furthermore, we have been able to establish that, by fixing the upper truncation at $m_{max}=m_{min}'=7.67$, the truncated power-law hypothesis cannot be rejected ( see Table \ref{tab:universaltable}). Consequently, if one wants to fit a power-law PDF with a unique exponent, one has to exclude all those earthquakes with $m \geq m_{max}= 7.67$. For the CMT catalog, we fix the upper cut-off $x_{max} =10^{\frac{3}{2}m_{max}+9.1}$ and we work by following the same procedure we have been using so far. The rest of catalogs can be safely fitted by untruncated power-law PDFs because the CV-test does not reject the hypothesis of a unique power-law tail and the magnitudes which are studied are considerably smaller than those in the CMT catalog. 

Thus, we consider two untruncated power-law distributions and a third one which is truncated for the CMT catalog. For each decade, we sample 5 values of $x_{min}^{(i)}$ equally spaced in logarithmic scale, and all the possible combinations of cut-offs $x_{min}^{(1)}$, $x_{min}^{(2)}$, $x_{min}^{(3)}$ are checked for a fixed upper-truncation $x_{max}^{(3)}$. The labels (1),(2) and (3) correspond to the catalogs YSH B, YSH A and CMT respectively.

\begin{table*}[htbp]
\begin{tabular}{|ll|r|r|r|l|l|l|l|}
\hline
 & $p_{c}=0.05$  & \multicolumn{1}{l|}{\textbf{$n$}} & \multicolumn{1}{l|}{\textbf{$m_{min} $}} &  \multicolumn{1}{l|}{\textbf{$x_{min}$ (Nm)}} &  \multicolumn{1}{l|}{\textbf{OM}}  &  \multicolumn{1}{l|}{\textbf{$\hat{b}$-value}}  &   \multicolumn{1}{l|}{\textbf{$\hat{\gamma}$}} & \multicolumn{1}{l|}{\textbf{$p_{fit}$}} \\ \hline
  & \textbf{(1) YHS B} & 3412 & 1.93 & $10^{12}$ & 5.18 & $0.99(1)$  & $1.633(7)$ & $0.072(8)$   \\ \hline 
 & \textbf{(2) YHS A} & 3500 & 3.27 &  $10^{14}$  & 5.90 &  $ 0.99(2)$ & $1.66(1) $ & $0.089(9)$  \\ \hline
 & \textbf{(3) CMT} &  19003 & 5.40 & $1.58 \times 10^{17}$ & 3.40 & $ 0.98(8)$ & $1.655(5) $ & $0.26(1)$ \\ \hline\hline
  &  \textbf{Model} $\alpha$ & \multicolumn{1}{l|}{\textbf{$\mathcal{N}$}} & \multicolumn{1}{l|}{} &  \multicolumn{1}{l|}{} & \multicolumn{1}{l|}{$\sum$\textbf{OM}} & \multicolumn{1}{l|}{\textbf{$\hat{b}_{g}$-value}} &  \multicolumn{1}{l|}{\textbf{$\hat{\Gamma}$}} & \multicolumn{1}{l|}{$p_{fit}$} \\ \hline
  & \textbf{} & 25915 & 1.93 & $10^{12}$ & 14.48 & $0.991(6)$ &  $1.661(4)$  & $0.079(9)$ \\ \hline 
 \hline 
\hline
 &   $p_{c}=0.20$ & \multicolumn{1}{l|}{\textbf{$n$}} & \multicolumn{1}{l|}{\textbf{$m_{min} $}} &  \multicolumn{1}{l|}{\textbf{$x_{min}$ (Nm)}} &  \multicolumn{1}{l|}{\textbf{OM}}  &  \multicolumn{1}{l|}{\textbf{$\hat{b}$-value}}  &   \multicolumn{1}{l|}{\textbf{$\hat{\gamma}$}} & \multicolumn{1}{l|}{\textbf{$p_{fit}$}} \\ \hline
  & \textbf{(1) YHS B} & 3412 & 1.93 & $10^{12}$ & 5.18 & $0.99(1)$  & $1.633(7)$ & $0.072(8)$   \\ \hline 
 & \textbf{(2) YHS A} & 3500 & 3.27 &  $10^{14}$ &  5.90 &  $ 0.99(2)$ & $1.66(1) $ & $0.089(9)$  \\ \hline
 & \textbf{(3) CMT} &  10422 & 5.67 & $3.98 \times 10^{17}$ & 3 &  $1.00(1)$ & $1.663(7)$  & $0.62(2)$ \\ \hline\hline
  &  \textbf{Model} $\alpha$  & \multicolumn{1}{l|}{\textbf{$\mathcal{N}$}} & \multicolumn{1}{l|}{} & \multicolumn{1}{l|}{} &  \multicolumn{1}{l|}{$\sum$\textbf{OM}} &  \multicolumn{1}{l|}{\textbf{$\hat{b}_{g}$-value}} &  \multicolumn{1}{l|}{\textbf{$\hat{\Gamma}$}} & \multicolumn{1}{l|}{$p_{fit}$} \\ \hline
  & \textbf{} & 17334 & 1.93 & $10^{12}$ & 14.08 & $1.000(8)$ &  $1.667(5)$  & $0.326(5)$ \\ \hline   
\end{tabular}
\caption{Results of fitting the models $\alpha$ and $\beta$ to the earthquake catalogs when doing the goodness-of-fit test with the CKSD statistis and different values of $p_c$. If the goodness-of-fit test is performed with the KSDMD statistic, the same set of cut-offs as the fit done by using CKSD$_{0.20}$ is found (see Table \ref{tab:statistics}). Same symbols as in Table \ref{tab:universaltable}. The number of orders of magnitude $OM= \log_{10}\left( \frac{x_{max}^{(i)}}{x_{min}^{(i)}}  \right)$ covered by each fit as well as for model $\alpha$ is also shown. The value $x_{max}^{(i)}$ is replaced for $x_{top}^{(i)}$ for untruncated fits. Note that $7.67-1.93=5.74$ units in magnitude correspond to $8.6$ orders of magnitude in seismic moment. }
  \label{tab:globalfitting2}
\end{table*}

\begin{table*}[htbp]

\begin{tabular}{|ll|r|r|}
\hline
& \multicolumn{1}{l|}{} & \multicolumn{1}{l|}{\textbf{$D_{e}$}} & \multicolumn{1}{l|}{\textbf{$p_{fit}$}} \\ \hline
    & \textbf{KSDMD} & 0.018487 & $0.20(1)$    \\ \hline 
 & \textbf{CKSD$_{0.05}$} & 3.25082 & $0.079(9)$  \\ \hline
  & \textbf{CKSD$_{0.20}$} & 2.778202 & $0.326(5)$  \\ \hline
\end{tabular}
\caption{Results of fitting model $\alpha$ to all the catalogs when doing the goodness-of-fit test with the KSDMD and the CKSD statistics for $p_{c}=0.05$ (same cut-offs for all the catalogs) and for the set-of cut-offs for which the $p$-value is larger or equal than $p_{c}=0.20$ for the CKSD statistic. KSDMD has a unique fit because the fit already exceed $p_c=0.20$ and no valid fits are found with a smaller $p$-value.}
  \label{tab:statistics}
\end{table*}

\begin{figure*}[htpb]
\resizebox{\textwidth}{!}{\includegraphics[scale=0.70]{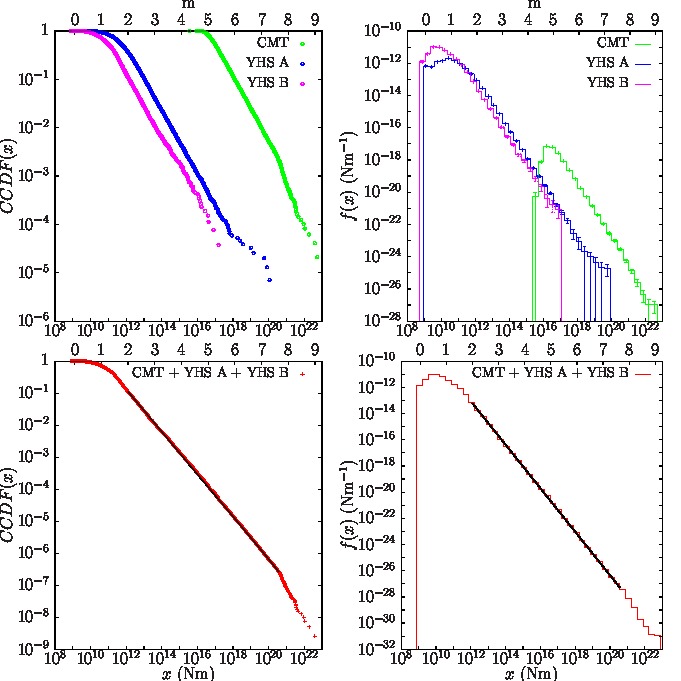}}
 \caption{Left panel: Estimated Complementary Cumulative Distribution Functions (CCDF) of the Gutenberg-Richter law for each catalog (top) and for the merged catalogs (bottom). Right panel: Estimated PDFs $f(x)$ of the Gutenberg-Richter law for each catalog (top) and for the merged catalogs (bottom). Merged histogram is plotted by following the procedure explained in Ref.~\cite{Navas2018}. Fits are represented by solid black lines. Top axis represents the same scale in moment magnitude.} \label{fig:GRGlobal} 
\end{figure*}

In Table \ref{tab:globalfitting2} we present the results of the global fit for models $\alpha$ and $\beta$. The same global fit is found independently from the choice of the statistic of the goodness-of-fit test. The values of the statistics as well as the resulting $p$-values are shown in Table \ref{tab:statistics}. It can be observed that, for this particular set of cut-offs, the CKSD test has a smaller $p$-value and can be considered to be a more strict statistic with respect the KSDMD. No fit with a smaller $p$-value has been found for the KSDMD statistic, and the fit shown in Table \ref{tab:globalfitting2} has a $p$-value that exceeds both $p_c=0.05$ and $p_c=0.20$. A $b$-value very close to one holds for more than 8 orders of magnitude in seismic moment from $m_{min}=1.93$ to $m_{max}=7.67$. The value of the global exponent is approximately in agreement with the harmonic mean of the particular exponents of the GR-law for each catalog \cite{Navas2018,kamer2015}. Due to the upper-truncation, the value of the global exponent is not exactly the same as the value of the harmonic mean of the particular exponents. The results do not show remarkable differences if the critical $p$-value is set to $p_c=0.20$ (see Table \ref{tab:globalfitting2}). We consider that the definitive fit is the one whose goodness-of-fit test has been performed with a threshold value of $p_c=0.20$.

\subsection{Universality of Earthquakes and Charcoal labquakes}
\label{sec:earthquake+charcoal}
\begin{table*}[htbp]
\begin{tabular}{|ll|r|r|r|l|l|l|l|}
\hline
 & \textbf{$p_c =0.05$} & \multicolumn{1}{l|}{\textbf{$n$}} & \multicolumn{1}{l|}{\textbf{$m_{min} $}} &  \multicolumn{1}{l|}{\textbf{$x_{min}$ (Nm)}} &  \multicolumn{1}{l|}{\textbf{OM}}  &  \multicolumn{1}{l|}{$\hat{b}$-value}  &   \multicolumn{1}{l|}{\textbf{$\hat{\gamma}$}} & \multicolumn{1}{l|}{\textbf{$p_{fit}$}} \\ \hline
 & \textbf{(0) Charcoal} & 15906 & - & $6.31\times 10^{-18}$ & 6.47 & $0.988(8)$ &   $1.658(5)$  &  $0.15(1)$ \\ \hline 
 & \textbf{(1) YHS B} & 1353 & 2.33 & $9.99 \times 10^{7}$ & 4.58 & $0.98(3)$ &  $ 1.66(2) $  & $0.10(1)$   \\ \hline 
 & \textbf{(2) YHS A} & 234 & 4.47 & $1.59 \times 10^{11}$ & 4.10 & $0.98(6)$ &   $1.65(4)$  & $0.62(2)$ \\ \hline
 & \textbf{(3) CMT} &  7689 & 5.80 & $1.59 \times 10^{13}$ & 2.80 & $1.00(1)$  & $1.667(9)$ & $0.393(5)$  \\ \hline\hline
 & \textbf{Model} $\alpha$  & \multicolumn{1}{l|}{\textbf{$\mathcal{N}$}} & \multicolumn{1}{l|}{} &  \multicolumn{1}{l|}{} &  \multicolumn{1}{l|}{$\sum$\textbf{OM}} &   \multicolumn{1}{l|}{\textbf{$\hat{b}_{g}$-value}} &  \multicolumn{1}{l|}{\textbf{$\hat{\Gamma}$}} & \multicolumn{1}{l|}{$p_{fit}$} \\ \hline
 & \textbf{CKSD $D_{e}=3.589973$} & 25182 &  & $6.3 \times 10^{-18}$& 17.95 & $1.003(6)$ & $1.669(4)$  & $0.057(7)$ \\ \hline 
\hline \\
\hline
   & \textbf{$p_c = 0.20$} & \multicolumn{1}{l|}{\textbf{$n$}} & \multicolumn{1}{l|}{\textbf{$m_{min} $}} &  \multicolumn{1}{l|}{\textbf{$x_{min}$ (Nm)}} & \multicolumn{1}{l|}{\textbf{OM}} & \multicolumn{1}{l|}{$\hat{b}$-value}  & \multicolumn{1}{l|}{\textbf{$\hat{\gamma}$}} & \multicolumn{1}{l|}{\textbf{$p_{fit}$}} \\ \hline
 & \textbf{(0) Charcoal} & 3555 & - & $6.31 \times 10^{-17}$  & 5.47 & $1.04(2)$ & $1.69(1)$ & $0.88(1)$ \\ \hline 
 & \textbf{(1) YHS B} & 1007 & 2.47 & $1.59 \times 10^{8}$ & 4.38 & $1.66(2)$ & $0.99(3)$  & $0.058(7)$   \\ \hline 
 & \textbf{(2) YHS A} & 234 & 4.47 & $1.59 \times 10^{11}$ & 4.10 & $0.98(6)$ & $1.65(4)$  & $0.62(2)$ \\ \hline
 & \textbf{(3) CMT} &  3014 & 6.20 & $6.33 \times 10^{13}$ & 2.20 & $1.00(2)$  & $1.67(2)$ & $0.59(2)$  \\ \hline\hline
 & \textbf{Model} $\alpha$  & \multicolumn{1}{l|}{\textbf{$\mathcal{N}$}} & \multicolumn{1}{l|}{} &  \multicolumn{1}{l|}{}&  \multicolumn{1}{l|}{$\sum$\textbf{OM}} & \multicolumn{1}{l|}{\textbf{$\hat{b}_{g}$-value}} &  \multicolumn{1}{l|}{\textbf{$\hat{\Gamma}$}} & \multicolumn{1}{l|}{$p_{fit}$} \\ \hline
 & \textbf{CKSD $D_{e}=2.998867$} & 7810 & & $6.3 \times 10^{-17}$ & 16.15 & $1.03(1)$ & $1.688(8)$  & $0.21(1)$ \\ \hline 
\end{tabular}
\caption{ Results of fitting models $\alpha$ and $\beta$ to the charcoal and earthquake datasets for two different values of $p_c$. Same notation as in previous tables. }
  \label{tab:universaltable2}
\end{table*}

\begin{figure*}[htpb]
\includegraphics[width=\textwidth]{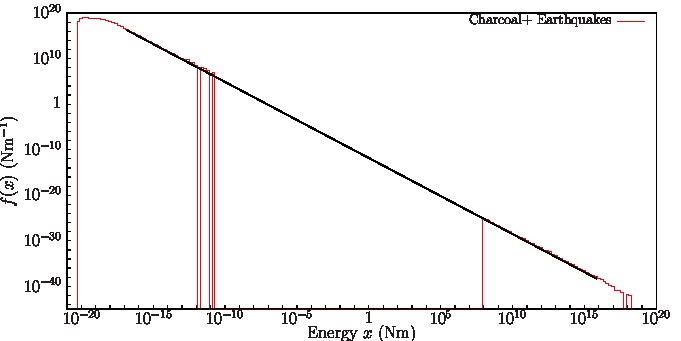} 
 \caption{Estimated PDF $f(x)$ of the Gutenberg-Richter law for the merged earthquake and charcoal labquake catalogs. The fit is represented by a solid black line. The methodology to merge the three earthquake catalogs is the same as in the rest of plots whereas the addition of the charcoal catalog to this fit has been done \textit{ad hoc} by conveniently rescaling both parts (those corresponding to charcoal and earthquakes respectively).} \label{fig:GRUniversal} 
\end{figure*}

Motivated by the fact that the power-law exponents of earthquake catalogs and charcoal labquakes are very similar (see Table \ref{tab:universaltable}), we apply this methodology by adding also the charcoal catalog.

At this step it is important to stress the fact that the seismic moment does not correspond with the radiated energy $E_{r}$ by the earthquake, which would be the reasonable energy to compare with the AE energy. 
Whether the ratio of seismically radiated energy over the seismic moment is independent on the moment magnitude is still an unsolved question \cite{KB04}. A constant ratio would imply that the static stress drop is constant for all the earthquakes. This ratio may depend on different earthquake parameters such as moment magnitude and the depth of the source \cite{BrodskyKanamori2001,VassiliouKanamori1982,IzutaniKanamori2001,KanamoriHeaton2000}. On the contrary, the seismically radiated energy is in some cases underestimated such as the ratio may be considered as constant \cite{IdeBeroza2001,Wang2014,BilekLayRuff2004,ZolloOreficeConvertito2014}. For our study, we are going to consider this ratio as constant so that the values of the seismic moment should just be multiplied by an unique factor. The value of this unique factor is $\frac{E_r}{M}=10^{-4.6}$ Ref.~\cite{Bormann2015}, where $M$ is the moment magnitude (previously called $x$). In this case, $x$ corresponds to the energy radiated in seismic waves by earthquakes $E_{r}$ and the AE energy.

It can be shown that, for both models $\alpha$ and $\beta$, multiplying the variable by a constant factor only introduces a constant term in the log-likelihood that does not change neither the maximum nor the difference of the log-likelihoods. As the CKSD statistic is a weighted average of the particular KS distances of each dataset, it does not change neither. Therefore, the results shown in Table \ref{tab:universaltable} would not change except for the values of the cut-offs.

The CV-test does not reject the hypothesis of a unique power-law tail for the charcoal catalog and an untruncated power-law model is considered for this catalog.
For each decade, 5 different values of $x_{min}^{(i)}$ equally spaced in logarithmic scale, for a fixed upper-truncation $x_{max}^{(3)}$ are checked and all the possible combinations of cut-offs $x_{min}^{(0)}$, $x_{min}^{(1)}$, $x_{min}^{(2)}$, $x_{min}^{(3)}$ are checked for a fixed upper-truncation $x_{max}^{(3)}$. The labels (0),(1),(2) and (3) correspond to the catalogs of the charcoal experiment, YSH B, YSH A and CMT respectively.
In Table \ref{tab:universaltable2} we present the results of the global fit for the CKSD statistic. In this case, not all the catalogues overlap each other and the CKSD statistic is the only one that can be used for the goodness-of-fit test. 
The value of the global exponent is approximately in agreement with the harmonic mean of the particular exponents of the GR-law for each catalog \cite{Navas2018}.  The results do not show remarkable differences if the critical $p$-value is set to $p_c=0.20$ ( see Table \ref{tab:universaltable2}). We consider that the definitive fit is the one whose goodness-of-fit test has been performed with a threshold value of $p_c=0.20$.

\subsection{Vycor labquakes}
\label{sec:vycor}
The values of the particular exponents $\gamma_i$ of Vycor labquake catalogs differ remarkably from those found for earthquakes and charcoal labquakes \cite{Navas2018}. No combination of the cut-offs has lead to a good fit if Vycor labquakes ar merged with charcoal labquakes or earthquakes. Therefore, Vycor labquakes can be considered in a different universality class.  

Due to the limitations due to saturation effects for high energy AE events, it is convenient to check whether it is necessary an upper truncation for the power-law regime. The residual CV-test reveals the upper truncations of the experiments performed at $60$,$40$ and $0$ dB but it does not provide conclusive results for the experiment at $20$dB. In order to explore fewer combinations of cut-offs, thus reducing the computation time, we fix the upper truncations $x_{max}^{(0)}$, $x_{max}^{(1)}$ and $x_{max}^{(3)}$ whereas $x_{max}^{(2)}$ is considered as a free parameter. The labels (0),(1),(2) and (3) correspond to the experiments performed at 60, 40, 20 and 0 dB respectively.
Five intervals per decade equally spaced in logarithmic scale are sampled for each catalog and all the possible combinations of cut-offs $x_{min}^{(0)}$, $x_{min}^{(1)}$, $x_{min}^{(2)}$, $x_{max}^{(2)}$, $x_{min}^{(3)}$ are checked for fixed upper-truncations $x_{max}^{(0)}$, $x_{max}^{(1)}$ and $x_{max}^{(3)}$. 

In Table \ref{tab:globalvycor} we present the results of the global fits for Vycor catalogs for the KSDMD and the CKSD statistics. In both cases, the global exponents are very similar but the fit performed with the KSDMD statistic maximizes the sum of orders of magnitude $\sum_{i=1}^{n_{ds}} \log_{10} \left( \frac{x_{max}^{(i)}}{x_{min}^{(i)}} \right)$ and will constitute the definitive fit. The same combination of cut-offs would not be a good fit if the CKSD statistic is used instead. Although the CKSD statistic is more restrictive in this case, it is preferable to consider the maximum number of orders magnitude because it cannot be ensured that this statistic will always be more restrictive for different datasets. The definitive fit covers about nine orders of magnitude in energy with a number of data $\mathcal{N}=30005$ entering into the fit. The results do not differ remarkably if the critical $p$-value is set to $p_c=0.20$ (See Table \ref{tab:globalvycor}).

\begin{figure*}[htpb]
  \resizebox{\textwidth}{!}{\includegraphics[scale=0.70]{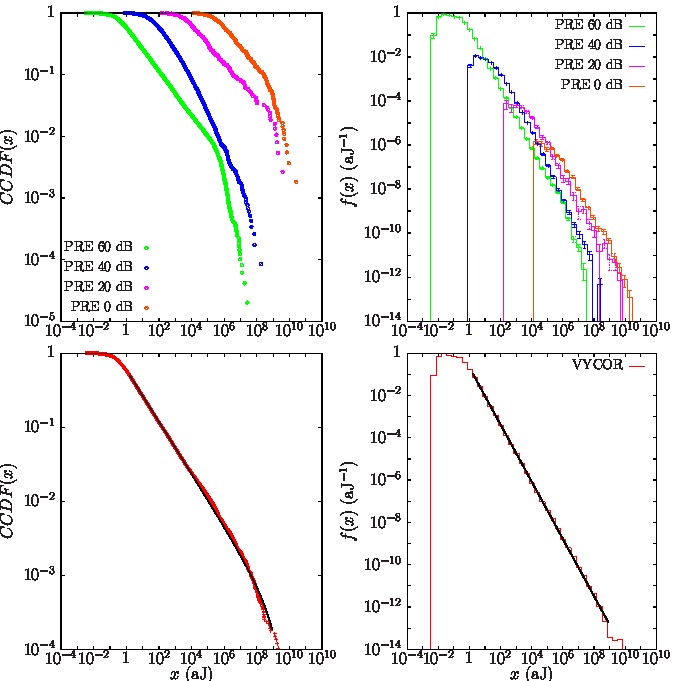}}
 \caption{Left panel: Estimated Complementary Cumulative Distribution Functions (CCDF) of the Gutenberg-Richter law for each Vycor catalog (top) and for the merged Vycor catalogs (bottom). Right panel: Estimated PDFs $f(x)$ of the Gutenberg-Richter law for each Vycor catalog (top) and for the merged catalogs (bottom). Merged histogram is plotted by following the procedure explained in Ref.~\cite{Navas2018}. Fits are represented by solid black lines. } \label{fig:vycor} 
\end{figure*}

\begin{table*}[htbp]
\begin{tabular}{|ll|r|r|r|r|r|r|r|r|r|}
\hline
 & \textbf{$p_{c}=0.20$}  & \multicolumn{1}{l|}{\textbf{$n$}} & \multicolumn{1}{l|}{\textbf{$x_{min}$ (aJ)}} &  \multicolumn{1}{l|}{\textbf{$x_{top}$ (aJ)}} & \multicolumn{1}{l|}{\textbf{$x_{max}$ (aJ)}} & \multicolumn{1}{l|}{\textbf{OM}} &  \multicolumn{1}{l|}{\textbf{$\hat{\gamma}$}} &  \multicolumn{1}{l|}{$p_{fit}$}  \\ \hline
  & \textbf{(0) PRE 60} & 24338  & 1.58 & $2.66 \times 10^7$ & $5.37 \times 10^5$ & 5.53 & $1.351(2)$ & $0.080(6)$  \\ \hline 
 & \textbf{(1) PRE 40} & 5083  & 158.489 & $1.82 \times 10^8$  & 15860 & 2.00 & $1.34(1)$ & $0.087(9)$ \\ \hline
 & \textbf{(2) PRE 20} & 263  & 6309.57 & $3.76 \times 10^9$ & $10^9$ & 5.20 & $1.30(2)$ & $0.67(1)$ \\ \hline
  & \textbf{(3) PRE 0} & 321  & $6.31 \times 10^5$ & $2.62 \times 10^{10}$ & $3.19 \times 10^{8}$ & 2.70 & $1.31(1)$ &  $0.25(1)$   \\ \hline\hline
  &  \textbf{Model} $\alpha$ & \multicolumn{1}{l|}{\textbf{$\mathcal{N}$}} & \multicolumn{1}{l|}{} &  \multicolumn{1}{l|}{} & \multicolumn{1}{l|}{} &  \multicolumn{1}{l|}{$\sum$ \textbf{OM}} &  \multicolumn{1}{l|}{\textbf{$\hat{\Gamma}$}} & \multicolumn{1}{l|}{} \\ \hline
  & KSDMD $D_{e}=0.005911$& 30005  & 1.58 & $2.62 \times 10^{10}$  & $10^9$  & 15.43 & $1.350(2)$  & $0.20(1)$  \\ \hline
\\ \hline\hline
 & \textbf{$p_{c}=0.05$}  & \multicolumn{1}{l|}{\textbf{$n$}} & \multicolumn{1}{l|}{\textbf{$x_{min}$ (aJ)}} &  \multicolumn{1}{l|}{\textbf{$x_{top}$ (aJ)}} & \multicolumn{1}{l|}{\textbf{$x_{max}$ (aJ)}} & \multicolumn{1}{l|}{\textbf{OM}} &  \multicolumn{1}{l|}{\textbf{$\hat{\gamma}$}} &  \multicolumn{1}{l|}{$p_{fit}$} \\ \hline
  & \textbf{(0) PRE 60} & 24338 & 1.58 & $2.66 \times 10^7$ & $5.37 \times 10^5$ & 5.53 & $1.351(2)$ & $0.080(6)$ \\ \hline 
 & \textbf{(1) PRE 40} & 4179 & 251.189 &  $1.82 \times 10^8$ & 15860 & 1.80 & $1.35(1)$ & $0.19(1)$  \\ \hline
 & \textbf{(2) PRE 20} &  181 &  $2.51 \times 10^4$ & $3.76 \times 10^9$ & $2.51 \times 10^9$ & 5 & $1.29(3)$ &  $0.18(1)$  \\ \hline
  & \textbf{(3) PRE 0} & 198  & $2.51 \times 10^6$ & $2.62 \times 10^{10}$ & $3.19 \times 10^{8}$ & 2.10 & $1.36(6)$ & $0.36(2)$  \\ \hline\hline
  &  \textbf{Model} $\alpha$ & \multicolumn{1}{l|}{\textbf{$\mathcal{N}$}} & \multicolumn{1}{l|}{} &  \multicolumn{1}{l|}{} & \multicolumn{1}{l|}{} &  \multicolumn{1}{l|}{$\sum$ \textbf{OM}} &  \multicolumn{1}{l|}{\textbf{$\hat{\Gamma}$}} & \multicolumn{1}{l|}{} \\ \hline
  & CKSD $D_{e}=3.362608$ & 28896 & 1.58 & $2.62 \times 10^{10}$ & $2.51 \times 10^9$ & 14.43 & $1.351(2)$ & $ 0.052(7)$ \\ \hline 
  \\ \hline\hline
 & $p_c =0.20$ & \multicolumn{1}{l|}{\textbf{$n$}} & \multicolumn{1}{l|}{\textbf{$x_{min}$ (aJ)}} &  \multicolumn{1}{l|}{\textbf{$x_{top}$ (aJ)}} & \multicolumn{1}{l|}{\textbf{$x_{max}$ (aJ)}} & \multicolumn{1}{l|}{\textbf{OM}} &  \multicolumn{1}{l|}{\textbf{$\hat{\gamma}$}} &  \multicolumn{1}{l|}{$p_{fit}$} \\ \hline
  & \textbf{(0) PRE 60} & 20778 & 2.51 & $2.66 \times 10^7$ & $5.37 \times 10^5$ & 5.33 & $1.354(3)$ & $0.51(2)$ \\ \hline 
 & \textbf{(1) PRE 40} & 3404 & 398.107 &  $1.82 \times 10^8$ & 15860 & 1.60 & $1.36(2)$ & $0.36(2)$  \\ \hline
 & \textbf{(2) PRE 20} &  155 &  $3.98 \times 10^4$ & $3.76 \times 10^9$ & $10^9$ & 4.40 & $1.32(4)$ &  $0.20(1)$  \\ \hline
  & \textbf{(3) PRE 0} & 159  & $3.98 \times 10^6$ & $2.62 \times 10^{10}$ & $3.19 \times 10^{8}$ & 1.90 & $1.34(7)$ & $0.29(1)$ \\ \hline\hline
  & \textbf{Model} $\alpha$ & \multicolumn{1}{l|}{\textbf{$\mathcal{N}$}} & \multicolumn{1}{l|}{} &  \multicolumn{1}{l|}{} & \multicolumn{1}{l|}{} &  \multicolumn{1}{l|}{$\sum$ \textbf{OM}} &  \multicolumn{1}{l|}{\textbf{$\hat{\Gamma}$}} & \multicolumn{1}{l|}{} \\ \hline
  & \textbf{CKSD $D_{e}=2.586434$} & 24496 & 2.51 & $2.62 \times 10^{10}$ & $10^9$  & 13.23 & $1.354(2)$ & $ 0.20(1)$ \\ \hline 
  
\end{tabular}
\caption{Results of fitting models $\alpha$ and $\beta$ to the Vycor Labquake catalogs when doing the goodness-of-fit test with the KSDMD and the CKSD statistics and different values of $p_c$. KSDMD has a unique fit because the fit already exceed $p_c=0.20$ and no valid fits are found with a smaller $p$-value. Same notation as in previous tables. }
  \label{tab:globalvycor}

\end{table*}

\section{Conclusions}
\label{sec:conclusion}

We have presented a statistical procedure to merge different datasets in order to validate the existence of universal power-law exponents across different scales or phenomena. This methodology can be useful in the study of different complex systems.
In this work, the methodology has been applied to the Gutenberg-Ricther law for earthquakes and labquakes. By merging earthquake catalogs, a global power-law with a global exponent $\Gamma=1.667$ holds for more than 8 orders of magnitude in seismic moment (from $m_{min}=1.93$ to $m_{max}=7.67$ in moment magnitude). To our knowledge, this is the broadest fitting range that has been found for the Gutenberg-Richter law for earthquakes with a unique value of the exponent \cite{Kaganbeta}. 
There are catalogs of tiny mining-induced earthquakes which exhibit a much smaller completeness magnitude \cite{Jaguars2011} than the ones of natural seismicity used in this work. They were not considered here because they are not currently public and show $b$-values significantly different \cite{Jaguars2010} form the one found here, which would result in non-acceptable fits when merging them with the rest of catalogs, possibly pointing to a different universality class. Earthquake catalogs have been also merged with a charcoal labquake catalog with a global power-law exponent $\Gamma=1.688$ suggesting that these different systems might be classified into the same universality class. A previous methodology for merging datasets was not able to find a good fit for the Gutenberg-Richter law for the four Vycor Labquake catalogs \cite{Navas2018}. The previous procedure did not take into account the fact that the cut-offs of the merged catalogs could be different to those that limited the power-law regime for the individual catalogs. With the methodology exposed in this paper, which allows to consider different cut-offs, these Vycor labquake catalogs have been merged and a GR law with exponent $\Gamma=1.35$ for the energy AE events has been found to hold for nine orders of magnitude. The labquakes in this material have been found to belong to a different universality class than earthquakes and charcoal labquakes.


\section{Appendix}

\subsection{Appendix A: Likelihood Ratio Test }
\label{sec:appb}
The likelihood ratio test (LRT) is a statistical procedure used to check whether it is worth using a statistical model $\alpha$ (characterized by a set of $n_{\alpha}$ parameters) which is nested into a more general statistical model $\beta$ (characterized by a set of $n_{\beta}$ parameters). The null hypothesis of this test states that the simpler model $\alpha$ is good enough to describe the data and a more general model $\beta$ does not provide more information. In order to perform this test, the statistic which is used is:
\begin{equation}
\mathcal{R}= \log \left( \frac{\hat{\mathcal{L}}_{\beta}}{\hat{\mathcal{L}}_{\alpha}}  \right) =\log \hat{\mathcal{L}}_{\beta} - \log \hat{\mathcal{L}}_{\alpha}
\end{equation}
Where $\log$ is the natural logarithm and the hats denote that the likelihoods $\mathcal{L}_{\alpha}$ and $\mathcal{L}_{\beta}$ are evaluated at those values of the set of parameters for which they reach their maximum values.

Although not all the regularity conditions exposed in Ref.~\cite{pawitan2013all} are satisfied when dealing with power-law PDFs, it has been numerically tested that the statistic $2\mathcal{R}$ follows a chi-squared distribution with $n_{\beta}-n_{\alpha} >0$ degrees of freedom, at least for the largest percentiles \cite{Serra2017}. Once the significance level of the test has been fixed, one is able to determine the critical value $2\mathcal{R}_c$. If the empirical value of the statistic is below this threshold, then one can consider that there are not enough statistical evidences to say that the more complex model $\beta$ is needed in order to describe the data. In this situation, due to its simplicity, model $\alpha$ is preferable to fit data. This procedure can be naively understood as a statistical version of the Occam's razor criterion.

 Note that the rejection of the null hypothesis does not imply the rejection of model $\alpha$ as a fit to data, on the contrary, the ``acceptance" of model $\alpha$ does not imply that it is a good fit to the data. The test is just a relative comparison between two models. 

We present the expressions that are necessary to compute the statistic of the likelihood ratio test in this work.
The log-likelihood for the untruncated PDF in Eq. (\ref{eq:notrunc}) is written:
\begin{equation}
\log \mathcal{L}_{untrunc} \left( x; x_{min}, \gamma   \right) = n \log \left( \frac{\gamma-1}{x_{min}^{1-\gamma} }  \right)-\gamma \sum_{j=1}^{n} \log x_{j}
\label{eq:likepl}
\end{equation}
whereas the log-likelihood for the truncated PDF in Eq. (\ref{eq:trunc}) is written:
\begin{equation}
\begin{split}
\log \mathcal{L}_{trunc} \left( x; x_{min}, x_{max}, \gamma   \right) = & n \log \left( \frac{1-\gamma}{x_{max}^{1-\gamma} - x_{min}^{1-\gamma} }  \right)\\
& -\gamma \sum_{j=1}^{n} \log x_{j}
\end{split}
\label{eq:likepltrunc}
\end{equation}

\subsection{Appendix B: Residual Coefficient of Variation Test}
\label{sec:appa}

A useful statistical tool to check whether an untruncated power-law distribution is a good model to fit data in a certain range is the so called Residual Coefficient of Variation (CV) test \cite{Malevergne2011}. 

Let us suppose that we have a variable $\mathcal{X}$ that acquires $N$ values $\lbrace x \rbrace$. We define $x_{(j)}$ as the $j$-th value of the variable when it is sorted in ascending order: $x_{(1)} \leq x_{(2)} \leq ... \leq x_{(j)} \leq ... \leq x_{(N)}$. 
The null hypothesis of the test states that there exists a power-law tail given by $x > x_{(k)}$. 
The test is based on the idea that for an untruncated power-law distribution the logarithmic CV is close to 1:
\begin{equation}
CV_{l} = \frac{s_{l}}{m_{l}}
\end{equation}
where $m_l$ corresponds to the mean of the logarithm of the rescaled variable $l=\log(x/x_{(k)})$:
\begin{equation}
m_{l} = \frac{1}{N-k} \sum_{j=k+1}^{N} \log \left( \frac{x_{(j)}}{x_{(k)}}  \right)
\end{equation}
and $s^{2}_l$ is the unbiased variance:
\begin{equation}
s_{l}^2 = \frac{1}{N-k-1} \sum_{j=k+1}^{N} \left( \log \left( \frac{x_{(j)}}{x_{(k)}}  \right)-m_l  \right)^2
\end{equation}
In order to check whether the value of the residual CV is close to one or not for a particular number of remaining data, one simulates many samples power-law distributed with the same number of data $N-k$ in order to extract a distribution for $CV_{l}$. It is important to remark that the distribution of the statistic does not depend on the power-law exponent and, therefore, it is not necessary to estimate it previously. Once one determines the level of significance for the test (in this case $0.05$), one can obtain the upper and lower critical values of the statistic by checking the percentiles of the distribution of simulated values (percentiles $2.5$ and $97.5$ in our case). If the empirical value of $CV_{l}$ lies between these two critical values, one cannot reject the null hypothesis of power-law tail. Otherwise, if the empirical value is below percentile 2.5, the power-law is rejected in favour of a truncated log-normal; if the empirical value is above percentile 97.5 the power-law is rejected but there is no alternative. One proceeds by computing the residual $CV_{l}$ for increasing values of the index $k$, thus analysing the remaining values in the tail of distribution.



\subsection{Appendix C: Goodness-of-fit test}
\label{sec:appd}
Determining whether the null hypothesis of considering a global exponent $\Gamma$ is compatible with the values of the particular fits presents some differences with the standard Kolmogorov-Smirnov (KS) goodness-of-fit test presented in Ref.~\cite{Deluca2013}. In this appendix we expose the two different statistics for the goodness-of-fit test that have been used for the global fit: the first statistic is essentially an adaptation of the standard KS test for the merged case whereas the second method uses a statistic which is a composition of KS distances.
When the value of the global exponent $\Gamma$ has been found through MLE, one can understand that each dataset contributes to the global PDF with a global exponent $\Gamma$ in their particular ranges $\left[ x_{min}^{(i)} ,x_{max}^{(i)}  \right]$ ($i=1,...,n_{\rm cat}$). The global distribution can be understood as a global power-law PDF with exponent $\Gamma$ ranging from $X_{min}=\min \lbrace x_{min}^{(i)} \rbrace$ to $X_{max}=\max \lbrace x_{max}^{(i)} \rbrace$.
In this situation, we need to redefine the KS distance in this case where we have merged several datasets. 
\begin{itemize}
\item \textbf{Kolmogorov-Smirnov Distance of the Merged Datasets (KSDMD):} This statistic can be used as long as datasets overlap each other. 
Datasets can be merged by pairs by giving a certain weight to each point from data. 
Depending on which overlapping or non-overlapping region a point from our data comes from, each point will have a certain weight $\omega_{j}$ ($j=1,...,\mathcal{N}$). For more details about the expression of these weights see Ref.~\cite{Navas2018}. By sorting data in ascending order, it is easy to construct the empirical Cumulative Distribution Function (CDF) for the merged datasets by using the following expression:
\begin{equation}
CDF_e (x_{(k)}) = \frac{\sum_{j=1}^{k} \omega_j }{\sum_{j=1}^{\mathcal{N}}\omega_{j}},
\end{equation}
with $k=1,...,\mathcal{N}$ and the sub-index $e$ refers to the empirical CDF. Note that, for a standard situation in which datasets are not merged, all the weights would be one \cite{Deluca2013}. Note also that the Complementary Cumulative Distribution Function (CCDF) is $CCDF(x_{(k)})=1-CDF(x_{(k)})$.
Once we have constructed the merged $CDF_{e}(x)$, we can easily compute the KS distance by:
\begin{equation}
D_{e}^{(KSDMD)} = \max_{X_{min} \leq x \leq X_{max}} \biggr\vert \left( \frac{x^{1-\Gamma}-X_{min}^{1-\Gamma}}{X_{max}^{1-\Gamma}-X_{min}^{1-\Gamma}}  \right) - CDF_{e}(x)   \biggr\vert
\end{equation}

\item \textbf{Composite KS Distance (CKSD):}
This second statistic can be computed independently on whether the datasets are overlapping each other or not. This statistic is constructed from the $n_{\rm cat}$ particular KS distances:
\begin{equation}
D_{e,i}= \max_{x_{min}^{(i)} \leq x \leq x_{max}^{(i)}} \biggr\vert   \left( \frac{x^{1-\Gamma}-x_{min}^{(i)1-\Gamma}}{x_{max}^{(i)1-\Gamma}-x_{min}^{(i)1-\Gamma}}  \right) - CDF_{e,i}\left(  x; x_{min}^{(i)},x_{max}^{(i)}  \right)   \biggr\vert,
\end{equation}
where the subindex $i$ refers to the $i$-th catalog and $CDF_{e}$ is the empirical CDF.
In order to compute the statistic, we perform the following summation:
\begin{equation}
D_{e}^{(CKSD)}=\sum_{i=1}^{\rm n_{\rm cat}} \sqrt{ n_{i}} D_{e,i},
\end{equation}
where the factors $\sqrt{ n_{i}}$ are due to the scaling of the KS distance with the number of data \cite{Press}.

\end{itemize}

Once the statistic is computed, one needs to determine whether it is big or small in relation to the one found for data sampled from a PDF with the same parameters $\lbrace x_{min}^{(i)} \rbrace$,$\lbrace x_{max}^{(i)} \rbrace$, $\Gamma$ and $\lbrace n_{i} \rbrace$.
Data is sampled from a power-law PDF in the range given by the $i$-th catalog with probability $q_{i}=n_{i}/\mathcal{N}$, where $\mathcal{N}=\sum_{i=1}^{\rm n_{ds}} n_{i}$. Note that the particular number in each simulated dataset is not necessarily the empirical one $n_{i}$ but the total number of data $\mathcal{N}$ is maintained. Hence, one needs a first random number to choose the dataset $i$ and therefore the range $\left[ x_{min}^{(i)},x_{max}^{(i)} \right]$ and a second one to generate the random truncated power-law number in that range with exponent $\hat{\Gamma}$ \cite{Deluca2013}. When $\mathcal{N}$ events have been generated according to this procedure, one finds the global exponent $\hat{\Gamma}_{\rm sim}$ by maximizing the global log-likelihood and computes either $D_{sim}^{(KSDMD)}$ or $D_{sim}^{(CKSD)}$ for the merged simulated datasets. 
By performing several realizations of the previous procedure, one can estimate the $p$-value of the fit by computing the fraction of simulated datasets where the simulated statistic is larger than the empirical one.

\section*{Acknowledgements}
The research leading to these results has received founding
from ``La Caixa'' Foundation. V. N. acknowledges financial
support from the Spanish Ministry of Economy and Competitiveness
(MINECO, Spain), through the ``Mar\'{\i}a de Maeztu'' Programme
for Units of Excellence in R \& D (Grant No. MDM-2014-0445).
We also acknowledge financial support from MINECO under 
Grants No. FIS2015-71851-P, FIS-PGC 2018-099629-B-100 and 
 MAT2016-75823-R, and
from Ag\`encia de Gesti\'o d'Ajuts Universitaris i de Recerca
(AGAUR) under Grant No. 2014SGR-1307 and Juan de la Cierva research contract FJCI-2016-
29307 hold by A.G.


\end{document}